\documentstyle[aps,prl,eqsecnum,preprint]{revtex}
\begin{document}
\title{On the confinement of spinons in the $CP^{M-1}$ model }
\author{Andrey V. Chubukov${}^{1,2}$, Oleg A. Starykh${}^3$~\cite{oas}}
\address{
${}^{1}$Department of Physics, University of Wisconsin, Madison, WI 53706\\
${}^{2}$P.L. Kapitza Institute for
Physical Problems, Moscow, Russia\\
${}^{3}$ Texas Center for Superconductivity, University of Houston,
Houston, TX 77204-5932}
\date{today}
\maketitle
\begin{abstract}
We use the $1/M$ expansion for the $CP^{M-1}$ model to
 study the long-distance behaviour of the staggered spin  susceptibility
in the commensurate, two-dimensional quantum antiferromagnet at
finite temperature.
 At $M=\infty$ this model possesses deconfined  spin-1/2 bosonic
spinons (Schwinger bosons), and the
 susceptibility has
a branch cut along the imaginary $k$ axis.
 We show that in all three
scaling regimes at finite $T$, the interaction
between spinons and gauge field fluctuations leads to divergent
$1/M$ corrections near the branch cut. We identify the most divergent
corrections to the susceptibility at each order in $1/M$
and explicitly show that the full static staggered susceptibility
has a number of simple poles rather than a branch cut.
We compare our results with the $1/N$ expansion for the $O(N)$ sigma-model.

PACS: 67.50-b, 67.70+n, 67.50Dg

\end{abstract}
\pacs{}

\section{Introduction}
\label{sec1}

Numerous studies of the two-dimensional (2D) quantum Heisenberg
antiferromagnet (QHAF) undertaken in the last few years
have  significantly improved our understanding of the behavior of these systems
at low temperatures~\cite{chakr,csy,css}. There is no ordered state
at any finite $T$ (and hence,
no phase transitions), but there are nevertheless three distinct low-$T$
regimes depending
on the relative values of the temperature and the coupling constant, $g$.
 These three regimes are~\cite{chakr}: {\it (i)}
renormalized-classical (RC) regime, where $T$ is smaller than the spin
stiffness in the ordered ground state,
$\varrho_s$;
{\it (ii)}
quantum-disordered (QD) regime, when $g$ is larger than the critical coupling
for the $T=0$ disordering transition,
 and $T$ is smaller than the gap,
$\Delta$, between the singlet ground state and the lowest excited
state with $S=1$; {\it (iii)}
 quantum-critical (QC) regime, which lies between the other
 two, and in which temperature is the largest infrared cutoff,
$k_B T \gg \varrho_s,
\Delta$.

It is very likely that the low-energy physics of a 2D QHAF, at least in the
vicinity of the disordering transition, is adequately
described by  the $O(3)$ nonlinear
sigma-model (see Sec~\ref{sec1}). Some information about the properties of this
model can be obtained from the Bethe-ansatz solution~\cite{pasha,hasen}, but
most of the thermodynamic properties have been studied using several
available perturbative techniques. The first perturbative approach to the
$O(3)$ sigma-model was
initiated many years ago by Polyakov~\cite{pol}. In this approach,
one  departs from  the ordered state
at $T=0$ , and  applies renormalization group theory which accounts for the
effects of classical fluctuations at small but finite $T$.
The expansion parameter for the RG studies is $T/2\pi \varrho_s$.
It has to be small, i.e., the system should be in the RC regime.
 The RG approach  allows one to obtain a scale
at which the fluctuation corrections to the spin-wave coupling constant
become comparable to its bare value.
 This scale is then identified
with the correlation length in a system.
Calculations  along these lines yielded
$\xi = A (\hbar c/2\pi \varrho_s) \exp(2\pi \varrho_s/k_B T)$~\cite{Brezin}
and $S(q) = N_0^2~(k_B T/2 \pi \varrho_s)^2 \xi^2 f(q\xi)$~\cite{chakr}.
 Here $A$ is a constant ($A \approx 0.34$
{}~\cite{hasen}), $S(q)$ is the static structure factor, $N_0$ is the staggered
magnetization, and $f(x)$
tends to a finite value at $x=0$.

Another widely used approach is the $1/N$ expansion for the $O(N)$ sigma-model.
The physical results are obtained in this approach
by extending the perturbation series in $1/N$ to $N=3$.
The advantage of the $1/N$ expansion
 is that it works in all three scaling regimes, and the
point of departure is a {\it disordered} state at any finite temperature.
The weak point of the theory is the absence of the physically motivated
 small parameter for the physical case $N=3$. In the RC regime, however, the
series of $1/N$ terms can be explicitly
summed up, and for the physical case of $N=3$ one
obtains {\it
exactly} the same results as in the RG approach.
At arbitrary $N$, one finds
$\xi = A_N (\hbar c/k_B T) ((N-2) k_B T /2\pi \varrho_s)^{1/(N-2)}
\exp(2\pi\varrho_s/((N-2) k_B T)$
 and $S(q) = N_0^2~(k_B T /2\pi \varrho_s)^{(N-1)/(N-2)}~
\xi^2 f_N (q\xi)$ ($A_3 \equiv A,~f_3(x) \equiv f(x)$).
 The results of $1/N$ expansion for all three scaling
regimes are collected in~\cite{csy}.

Finally, the third approach is based on the spinon representation for spin
operators (Schwinger boson theory). Read and Sachdev have shown
explicitly~\cite{rs} that  the low-energy limit of the Schwinger
boson theory is  described by
a $CP^{1}$ sigma-model for a two-component complex
unit field (see below). To obtain this model, one has to
express the unit vector field of the $O(3)$ sigma-model
as a bilinear
combination of two Bose fields ($CP^{1}$ representation),
\begin{equation}
n_a = z^{\dagger}_\alpha {\sigma}_{\alpha \beta}^a z_\beta,
\label{A1}
\end{equation}
where $\alpha, \beta$ are $SU(2)$ indexes, $a=x,y,z$, and $\sigma^a$ are the
Pauli matrices.
This representation also
introduces a $U(1)$ gauge degree of freedom,
  because ${\vec n}$ remains invariant
under the transformation of the bosonic fields,
$z_\alpha ({\bf r},t)\rightarrow
z_\alpha ({\bf r},t)e^{i\varphi({\bf r},t)}$. Each $z-$
field quantum carries $S=1/2$ and is therefore a bosonic spinon.
The condition ${\vec n}^2 =1$, however, imposes a local
constraint $z^{\dagger}_\alpha z_\alpha =1$, which implies that spinons can
appear only in pairs.
A mean-field version of the Schwinger-boson theory
has recently received a lot of attention~\cite{aa}.
In the mean-field approximation, one reduces the on-site
constraint to a constraint imposed on the averaged quantities, and
decouples the term that is quartic in $z$ in the  spin Hamiltonian.
{}From the solution of the self-consistent
equations  in the RC regime one then obtains the spin
correlation length, $\xi \sim (k_B T/\varrho_s) \exp(2\pi\varrho_s/k_B T)$,
and the static structure factor, $S(q) \sim (k_B T/2 \pi \varrho_s)^2
 \xi^2 {\tilde f}(q\xi)$, where
${\tilde f}(x)$ has the same asymptotic behavior as $f(x)$~\cite{aa}.
The temperature dependence of $S(q)$
and the exponent in the expression for the
correlation length are the same as in other approaches but
the correlation length aquires an extra power of $T$ in the prefactor.
Because of this incorrect prefactor in $\xi$, the validity of the
Schwinger-boson mean-field theory has been questioned~\cite{chakrev}.
More recently, however, the Schwinger-boson approach
to a 2D antiferromagnet has
been applied in a systematic way, by means of a controllable $1/M$
expansion~\cite{ital,me2}.
To generate this expansion,  each $z-$ field was assumed to have
 $M$ components rather than two (this changes the symmetry of the underlying
sigma-model to $CP^{M-1}$). The $1/M$ computations
have been performed
for field-theory~\cite{ital} and condensed-matter~\cite{me2} applications. It
has been shown that the extra power of $T$ in the prefactor is indeed an
artifact of the mean-field, $M=\infty$, approximation:
the correct prefactor is
$T^{1 - 2/M}$, and it reduces to a constant for the physical case
 of $M=2$.
At the same time, the $1/M$ corrections to the static structure factor do not
change the power of $T$, and the mean-field result for $S(q)$ is therefore
valid for all
$M$. Clearly then,
for physical spins,  the $n-$field and the Schwinger-boson approaches
 yield the same temperature dependence of
the observables in the RC regime, as they indeed should.
One can therefore
 safely use any of these perturbative techniques.
 Notice however, that the series of regular $1/M$ corrections are poorly
convergent while the regular $1/N$ corrections  are usually small.
This makes the Schwinger boson  approach
less reliable for practical purposes than the $1/N$ expansion for the $O(N)$
sigma-model.

There is, however, another discrepancy between the Schwinger-boson and the
$n-$field approaches, which in our opinion has not been fully clarified
in the literature.
Namely, in the $n-$field approach, the staggered static
spin susceptibility $\chi_s (k, \omega =0)$ is proportional to the static
$n-$ field propagator which has a simple pole
along the imaginary $k$ axis,  at $k= \pm  i \xi^{-1}$.
 The residue of the pole is finite at $N= \infty$, and remains finite
in the physical case of $N=3$.
This result is valid in all
three scaling regimes. On the other hand, in the Schwinger-boson formalism,
the staggered  spin susceptibility is a
convolution of two spinon propagators. At the mean-field level, spinons behave
as free particles, and elementary calculations show that
the staggered susceptibility
has only a branch cut singularity at $k = 2 i m_0$, where $m_0$ is the
mass (inverse correlation length) of a Schwinger boson. Since the behavior
near the singularity in $\chi(ik,0)$  determines the long-distance
properties of
the spin correlators, the difference in the type of the singularity in the two
models leads to  different predictions about the
long-distance behavior of the correlation function. Obviously, one of these
predictions must be wrong.

In this paper, we show that the branch cut behavior of $\chi(ik,0)$ is also an
artifact of the mean-field Schwinger-boson formalism.
 We will
see that $1/M$ corrections are divergent near the branch cut, and eventually
transform the branch cut into a simple pole. This phenomenon
is closely related to
the confinement of spinons in the $CP^{M-1}$ model in $2+0$ dimensions,
first studied by Witten~\cite{wit}.
He found that massless gauge fluctuations in the $CP^{M-1}$ model give rise
to a linear confining potential between spinons, and this yields a bound state
with a mass, $m$, which is a nonanalytical function of $1/M$: $m =
2m_0 (1 + O(1/M^{2/3}))$. This result was
reproduced in a number of more recent papers (for a review
see, e.g., Ref.\cite{ital2} and
references therein). However, to the best of our knowledge,
the effect of the $1/M$ corrections on the
staggered susceptibility has not been studied in detail. The results of
such study will be reported in this paper. Besides the RC regime, we will also
study staggered susceptibility in the QD and QC regimes.

Before we proceed to the description of our calculations, it is useful to
specify which $1/M$ corrections are essential to our analysis. The point is
that in the RC regime, there exists
a self-energy correction to the spinon propagator
of the form $1/M \log \log (k_B T/ m_0)$. Since $m_0$ is itself exponential
in $T$, the double logarithm in fact reduces to $1/M \log (\varrho_s /k_B T)$.
A series
of these logarithmic terms give rise to the above mentioned change in
the power of temperature in the preexponential factor in $\xi$
 from $T$ to $T^{1-2/M}$.
 Below we will assume that these corrections are {\it already}
included into the expression for the Schwinger boson mass.
 We will therefore consider only {\it regular} $1/M$ corrections
which, as we will show, are responsible for the confinement.

The paper is organized as follows.  In Sec.~\ref{sec2}, we will briefly review
the large $M$ expansion for the $CP^{M-1}$ sigma model and present $M=\infty$
results for the correlation length, spin susceptibility, and
gauge field propagator. In Sec.~\ref{sec3},
we consider the static staggered susceptibility in the RC regime. We first
compute the lowest-order $1/M$ corrections, select the most divergent ones,
and then sum up the ladder series of divergent $1/M$ terms by reducing the
problem to a Schrodinger equation. Discrete solutions of this Schrodinger
equation will correspond to the {\it poles} in the staggered susceptibility.
In Sec~\ref{sec4} and \ref{sec5} we report analogous calculations for QD and QC
phases, respectively. Finally, in Sec~\ref{sec6} we state our conclusions and
discuss open questions.

\section{The sigma-model}
\label{sec2}

Our starting point is the partition function for the
$O(3)$ nonlinear sigma model in Euclidian space
\begin{equation}
{\cal Z}=\int D{n_l} \delta(n^2_l - 1) exp \left\{-\frac{\varrho^0_s}{2\hbar}
\int_{0}
^{1/T} d{\tau} \int d^2 {\bf r} \left[\frac {1} {c^2_{0}}(\partial_{\tau}
{n_l})^2 + (\partial_i{n_l})^2\right]\right\} ,
\end{equation}
where $i=x,y$; $l=1,2,3$, and  ${\varrho}^{0}_s$
and $c_0$ are the bare
spin stiffness and
spin-wave velocity. For simplicity, throughout the paper we choose the units
where $\hbar =1$ and  $c_0=1$.
Vector ${\bf n}$ describes local staggered magnetization.
In the $CP^1$ representation (\ref{A1}),
 ${\cal Z}$ transforms into
\begin{equation}
{\cal Z}=\int D{\bar{z}} D{z} \delta(|z|^2 - 1)~ exp \left\{-
2{\varrho}^{0}_{s} \int_{0}^{1/T} d{\tau} \int d^2{\bf r} \left[
|\partial_{\mu}z|^2 - |\bar{z} \partial_{\mu}z|^2\right]\right\} .
\label{zzz}
\end{equation}
Here ${\mu}={\tau}, x, y.$
Introducing the Hubbard-Stratonovich vector field $A_{\mu}$ to decouple
the quartic term, we obtain
\begin{equation}
{\cal Z}=\int D{\bar{z}} D{z} D{A_{\mu}} \delta(|z|^2 - 1)~exp
\left\{-2{\varrho}^{0}_{s}~\int_{0}^{1/T} d{\tau}\int d^2{\bf r}
|(\partial_{\mu} - i A_{\mu})z|^2 \right\} .
\end{equation}
Now we generalize the doublet $z$ to the M-component complex vector,
rescale the $z$ field to $z\rightarrow \sqrt{M} z$, and
 introduce the coupling constant
$g=M/2{\varrho}^0_s$. Introducing then the constraint into the action in a
standard way, we
obtain for the partition function
\begin{equation}
{\cal Z}=\int D{\bar{z}} D{z} D{A_{\mu}} D{\lambda} \exp \left\{
-\frac{1}{g} \int_{0}^{1/T} d{\tau} \int d^2{\bf r}
\left[|(\partial_{\mu} -iA_{\mu})z|^2 + i\lambda(|z|^2-M)\right]
\right\} .
\label{D1}
\end{equation}
This is the partition function  for the $CP^{M-1}$ model \cite{adda}.
The action in (\ref{D1}) is quadratic in the $\bar{z}, z$ fields,
and we can therefore integrate
them out. This generates an effective action for the $A_{\mu}$ and
$\lambda$ fields, which contains $M$ only as a prefactor \cite{pol,adda}.
At large M, this effective action can be well approximated
by the quadratic fluctuations of $A_{\mu}$ and $\lambda$ around their
saddle-point values~\cite{pol}
\begin{equation}
 i<\lambda>=m^2 , <A_{\mu}>=0 ,
\end{equation}
The value of the spinon mass, $m$, can be obtained in the
$1/M$ expansion by solving
the constraint equation to any given order in 1/M. The
 nonzero solution
for $m$ implies that the rotational symmetry is unbroken.
A more detailed discussion of the derivation of the $CP^{M-1}$
model can be found in ref.\cite{rs}.

At $M=\infty$, the theory is particularly simple, because it describes
free massive spinons.
The spinon Green's function $G_0(\vec{p},{\omega}_n)$ is given by
\begin{equation}
G_0(\vec{p},{\omega}_n)=\frac{1}{p^2 + {\omega}_n^2 + m_0^2} ,
\end{equation}
and the constraint equation reads
\begin{equation}
T{\sum}_{{\omega}_n} \int \frac{d \vec{p}}{(2\pi)^2}G_0(\vec{p},{\omega}_n)
=\frac{1}{g} .
\end{equation}
Using the Pauli-Villars regularization of ultraviolet divergencies, we obtain
\begin{equation}
\frac{m_0}{4\pi} + \frac{k_B T}{2\pi}\log\left(1 - e^{-\frac{m_0}{k_B
T}}\right)
= 2\left(\frac{1}{g_c}
-\frac{1}{g}\right) ,
\label{mNinf}
\end{equation}
where $g_c=\frac{8\pi}{\Lambda}$, and $\Lambda$ is the ultraviolet regulator.
Depending on the values of $g/g_c$ and the temperature, the
solutions of (\ref{mNinf}) are~\cite{csy,aa}:
{\it (i)} RC regime ~($\varrho_s > k_B T$):
$m_0 = k_B T \exp-(4\pi {\varrho}_s/MT)$, where
${\varrho}_s = (M/2)~(1/g - 1/g_c)$ is the renormalized stiffness;
{\it (ii)}  QD regime~($m_0 > k_B T$):
$m_0 = \Delta +  O(\exp{-\Delta/T})$, where $\Delta = 8\pi (1/g_c - 1/g)$;
 {\it (iii)}  QC regime ($T > \varrho_s, \Delta$):
$m_0 = \Theta k_B T$, where $\Theta =2\log[(\sqrt{5} +1)/2]$.

\subsection{Staggered susceptibility}

The static staggered susceptibility is defined in continuum limit as
\begin{equation}
\chi(k,0) \delta_{ab} = \frac{a^2}{N_s}~\int_0^{1/k_B T}
 d\tau \int\frac{d^2 {\bf r}}{(2\pi)^2}
<S_a({\bf r},\tau) S_b(0,0)> \exp [-i({\vec k} + {\vec Q})\cdot {\vec r}] ,
\end{equation}
where ${\vec k}$ is a small momentum,
${\vec Q} = (\pi/a, \pi/a)$, and $N_s$ is
the number of spins in the system. As each spin is a byproduct of $z-$fields,
the physical susceptibility is related to the polarization operator of $z$,
$\Pi(k,0) =\Pi(k) = T\sum_{\omega}~\int (d^2 q/4\pi^2)
 G ({\vec q},\omega) G ({\vec k} + {\vec
q},\omega)$. In the RC and QC regimes, the relation between $\chi (k,0)$ and
$\Pi (k)$ can be obtained in the same way as in ~\cite{css}. We find
\begin{equation}
\chi (k,0)=\frac{N_0^2}{2 \varrho_s^2} \Pi(k),
\end{equation}
where $N_0$ and $\varrho_s$ are the fully renormalized on-site
magnetization and spin-stiffness at $T=0$. Notice that there is a factor of $2$
difference with the analogous expression for frustrated systems~\cite{css}.
In the QD regime, the rescaling factor between susceptibility and polarization
operator can be related
to the overall factor in the local susceptibility~\cite{csy}.
However, in this regime, we will only
obtain $\Pi (k)$ up to an overall factor,
so there is no need  to discuss the exact relation between
$\chi (k)$ and $\Pi (k)$ in the QD regime.

\subsection{Polarization operator}
\label{sec2b}

We now present the expressions for the
polarization operator at $M = \infty$. In the RC regime, the summation over
frequency reduces to the $\omega =0$ term, and one obtains
\begin{equation}
\Pi_0(k)= k_B T\int \frac{d^2 p}{4\pi^2}~
\frac{1}{(\vec{p} - \vec{k}/2)^2 + m_0^2}
{}~\frac{1}{(\vec{p} + \vec{k}/2)^2 + m_0^2}.
\label{B1}
\end{equation}
The momentum integration yields~\cite{csy,css,ital}
\begin{equation}
\Pi_0(k)= \frac{k_B T}{4\pi \delta \sqrt{\delta^2 - m^{2}_0}}~
\log \left[\frac{\sqrt{\delta^2 - m^{2}_0} + \delta}{ m_0}\right] ,
\label{B111}
\end{equation}
where we  introduced $\delta^2= k^2/4 +  m_0^2$. Clearly, $\Pi_0 (k)$ has a
branch cut singularity
 at $\delta =0$, i.e., at $k^2 = - 4 m^{2}_0$. Near the singularity, we obtain
\begin{equation}
\Pi_0(k)= \frac{k_B T}{8 m_0 \delta}.
\label{B2}
\end{equation}
 The behavior of $\Pi (k)$
near the branch cut determines the long-distance behavior of
the spin susceptibility in real space. Evaluating the Fourier transform of
(\ref{B2}), we obtain $\chi (r) \sim exp(-r/\xi)/r$, where
$\xi = 1/2 m_0$ at the mean-field level. At the same time,
the susceptibility in the
$O(N)$ sigma-model has a simple
pole at $k^2 = - \xi^{-2}$, and
its  long-distance behavior is $\chi (r) \sim
\exp(-r/\xi) /\sqrt{r}$.

For the QD regime, the leading term in $\Pi(k)$ is the $T=0$ piece.
The frequency
summation is then replaced by the integration, and, to leading order in
$\delta$, one obtains
\begin{equation}
\Pi_0(k)=\frac{1}{16 \pi m_0} \log(\frac{m_0}{\delta}).
\label{B22}
\end{equation}
We see that the polarization operator has only a weak logarithmical
singularity.
In real space, the singularity in $\Pi_0 (k)$ gives rise to
$\chi (r) \sim \exp(-2m_0 r)/r^2$, which is again different from the mean-field
result for the $n$-field model in this regime,
 $\chi (r) \sim \exp(-r/\xi)/r$.

Finally, for the QC regime, elementary considerations show that the
most singular behavior in $\Pi(k)$ comes from the $\omega=0$ term in the
summation over frequency, and hence $\Pi (k)$ is still given by (\ref{B2}).
 This result is intuitively obvious as the QC regime
is the interpolation regime between the classical and the quantum-disordered
 ones, and the
singularity in $\Pi(k)$ is much stronger in the classical regime.

\subsection{Gauge field propagator}

In the next sections, we will show that the mean-field $(M=\infty)$ behavior
of susceptibilities
changes drastically at finite $M$. The $1/M$ corrections to susceptibility
include the
propagators of the gauge field, $A$,  and the constraint field, $\lambda$.
We will see that to study confinement,
it is sufficient to know the propagators
of $\lambda$ and $A$  at distances much larger then the spin
correlation length. At $k \ll \xi^{-1}$,
the fluctuations of $\lambda$ and of the temporal component
of the gauge field are well screened~\cite{csy,me2}. However, there is no
screening for the spatial part
of the gauge field~\cite{adda,ital,me2}. The spatial component of the
gauge field propagator is given by
\begin{eqnarray}
{\Pi}_{\mu\nu}(q,\Omega)&=& 2{\delta}_{\mu\nu}k_B T
{\sum}_{\omega}\int\frac{d{\vec
{k}}}{(2\pi)^2}G_0(k,\omega)\nonumber\\
&& -k_B T{\sum}_{\omega}\int\frac{d{\vec{k}}}{(2\pi)^2}
G_0(k,\omega)G_0(\vec{k}+\vec{q},\omega + \Omega)(2k+q)_{\mu}(2k+q)_{\nu} ,
\end{eqnarray}
where $k_0=\omega, q_0 =\Omega$.
 For the RC and the QC regimes, $\xi^{-1} \leq k_B T$,
 and at $k \ll \xi^{-1}$ we
can also restrict to the  $\Omega =0$ component of $\Pi$.
The evaluation of integrals in the limit of $q \ll
m_0$ is then straightforward, and we find
\begin{equation}
{\Pi}_{\mu\nu}(q,0)=\left(\delta_{\mu\nu} -\frac{q_{\mu}q_{\nu}}{q^2}\right)
\frac{q^2}{12\pi} T\sum_n \frac{1}{\omega_n^2 +m_0^2}.
\end{equation}
Inverting now $\Pi_{\mu\nu}$ in the Coulomb gauge, we find for the
 gauge field propagator $D_{\mu\nu}$ at small momenta
\begin{equation}
D_{\mu\nu}(q)=\left(\delta_{\mu\nu} -\frac{q_{\mu}q_{\nu}}{q^2}\right)
D(q)=\left(\delta_{\mu\nu} -\frac{q_{\mu}q_{\nu}}{q^2}\right)
\frac{12\pi A}{q^2},
\label{gauge}
\end{equation}
where $\mu(\nu)=x,y$, and the values of
 $A$ are
$A=m_0^2/k_B T$ (RC regime), ~$A=2 \Theta k_B T /\sqrt 5$ (QC regime). In the
QD regime at $T=0$,
$\Omega$ is a continuous
variable, and $q, \Omega$ are the components of a $3D-momentum$.
We found that
eq. (\ref{gauge}) is still valid in this regime, $A = 2 m_0$, and
$\mu$ and $\nu$ are running over $x, y, ~\text{and}~ \tau$.

We now consider separately $1/M$ corrections in the three scaling regimes.

\section{renormalized-classical regime}
\label{sec3}

\subsection{First order in $1/M$}

As we said above, we only have to consider the $1/M$ corrections
associated with
the fluctuations of the gauge field. Corrections related to the
fluctuations of the constraint field renormalize the
spinon mass, but these fluctuations are screened at $q <2m_0$ and are therefore
irrelevant for the confinement.
The first-order $1/M$ corrections to the
polarization operator $\Pi (k)$ are shown in
Fig.~\ref{f_o_dia},
the propagators and vertex function
 which appear in the $1/M$ expansion are collected
in Fig~\ref{vertices}.
A simple power counting argument
shows that the  most singular corrections appear if
 we select a contribution proportional to the
{\it external} momentum, $k$, at each
interaction vertex with the gauge field.
 We then obtain near $k = \pm 2im_0$, and
 setting $G(\vec{p}) = G(\vec{p},\omega =0)$,
\begin{eqnarray}
\Pi_{1/M}(k)=&& \frac{1}{M}~
(k_B T)^2~\int \frac{d^2 p d^2 l}{16\pi^4}~G(\vec{p}-\vec{k}/2)
G(\vec{p}+\vec{k}/2)G(\vec{p}+\vec{l}+\vec{k}/2)\nonumber\\&&\times
\left(2G(\vec{p}+\vec{k}/2)
- G(\vec{p}+\vec{l}-\vec{k}/2)\right) k_{\mu}k_{\nu}
(\delta_{\mu\nu} - \frac{l_{\mu}l_{\nu}}{l^2}) D(l) ,
\end{eqnarray}
or, in explicit form,
\begin{eqnarray}
\Pi_{1/M}(k)=&& - \frac{4m^{2}_0}{M} (k_B T)^2
\int \frac{ d^2 p d^2 l}{16\pi^4} ~D(l) ~\frac{\sin^2\psi}
{(p^2 + \delta^2)^2 + 4m_0^2 p^2 \cos^2\varphi} \nonumber\\&&\times
\Big(2 \frac{(\vec{p}+\vec{l})^2 +\delta^2 - i2m_0(p\cos\varphi + l\cos\psi)}
{((\vec{p}+\vec{l})^2+\delta^2)^2 +4m_0^2(p\cos\varphi + l\cos\psi)^2}
\times \frac{p^2 +\delta^2- i2m_0 p\cos\varphi}
{(p^2 +\delta^2)^2 +4m_0^2\cos^2\varphi}
\nonumber\\&&-\frac{1}
{((\vec{p}+\vec{l})^2+\delta^2)^2 +4m_0^2(p\cos\varphi + l\cos\psi)^2} \Big),
\label{B3}
\end{eqnarray}
where $\varphi$ and $\psi$ are the angles between ${\vec p}$ and $\vec k$,
and $\vec l$ and $\vec k$, respectively.
The key point in the computation of
 $\Pi_{1/M}(k)$, as well as in other computations later in the paper, is that
for $\delta^2 \ll m_0^2$ (which is the only one we consider),
the angular integration is confined to a region
where $\varphi, \psi \approx \pm \pi/2$~\cite{comm}.
For practical purposes, it is convenient to restrict the
angular integration to the vicinity of $\pi/2$, but  extend the
integration over $p$ and $l$ from $-\infty$ to $+\infty$.
Rescaling the angular variables,
$\varphi \rightarrow \frac{\varphi}{2m_0p}, \psi \rightarrow \frac{\psi}
{2m_0l}$, we then obtain
\begin{eqnarray}
\Pi_{1/M}(k)=- \frac{1}{M}~ \left(\frac{k_B T}{4\pi^2}\right)^2
\int_{-\infty}^{\infty} dp
\int_{-\infty}^{\infty} dl \int_{-\infty}^{\infty}
d\varphi \int_{-\infty}^{\infty} d\psi \frac{1}{(p^2+\delta^2)^2 +
\varphi^2} \nonumber\\
\left(2\frac{(p - l)^2 +\delta^2 - i(\varphi+\psi)}
{((p - l)^2 +\delta^2)^2 +(\varphi+\psi)^2}~
\frac{p^2 +\delta^2-i\varphi}{(p^2+\delta^2)^2+\varphi^2} -
\frac{1}{((p - l)^2 +\delta^2)^2 +(\varphi+\psi)^2}\right) .
\label{B4}
\end{eqnarray}
The angular integration extends up to  $|\varphi|, |\psi| \sim
O(1/\delta)$. The integrals are convergent in the ultraviolet, so we can safely
set the limits of the angular integration equal  to infinity.
The integration is now trivial, and we find
\begin{equation}
\Pi_{1/M}(k)=- \frac{1}{M}~\left(\frac{k_B T}{4\pi}\right)^2
\int_{-\infty}^{\infty} dp\int_{-\infty}^{\infty}dl D(l)\frac{1}{p^2+
\delta^2}\left(\frac{1}{p^2+\delta^2} - \frac{1}{(p-l)^2+\delta^2}\right) .
\end{equation}
We emphasize at this point that
only the momenta perpendicular to the external momentum $\vec{k}$
contribute to the angular integration. This implies that
 the evaluation of the $1/M$ correction to the
polarization operator in $2+0$ dimensions in fact reduces to a
 one-dimensional problem. This dimensional transmutation will play a key
role in our analysis of the series of $1/M$ terms.
Finally, substituting $D (l)$ from (\ref{gauge}) and performing
the momentum integration, we obtain
\begin{equation}
\Pi_{1/M}(k)= -\frac{3 \pi m_0^2 T}{M (k^2+4m_0^2)^2}.
\label{B5}
\end{equation}
 Notice the absence of divergencies in the integration over $l$ - this is the
result of including both self-energy and vertex
corrections into $\Pi_{1/M}(k)$.

In obtaining (\ref{B5}), we implicitly assumed
that $\delta^2 = (k^2 + 4 m^2_0)/4 \geq 0$. Calculations for $\delta^2 \leq 0$
proceed  along the same lines,
and the result is
\begin{equation}
\Pi_{1/M}(k)= i \frac{12 m_0^2 T}{M (k^2 + 4m_0^2)^2},
{}~~ k^2 + 4 m^2_0 \rightarrow -0 .
\label{B55}
\end{equation}

The $1/M$ corrections to the polarization operator in the RC regime were
studied earlier by  Campostrini and Rossi
\cite{ital}. They numerically evaluated the leading singularity
in $\Pi_{1/N}(k)$ for $\delta^2 \rightarrow + 0$ and found the same result as
in (\ref{B5}), with the numerical prefactor $9.425$ which, as we show,
 is in fact exactly $3\pi$. However, their estimate for $\Pi_{1/M}(k)$
 for negative
$\delta^2$ is inconsistent with (\ref{B55}).

Let us now discuss the $1/M$ results.
Our first observation is that the
 $1/\delta^4$ divergence of $\Pi_{1/M}(k)$ is
 stronger than one could expect assuming that $\Pi (k)$ preserves its form
(\ref{B2}), and the two-spinon mass has a regular $1/M$ expansion
 (the latter would correspond to $\Pi_{1/M}(k) \sim
O(\delta^{-3})$). A crude estimate that provides insight into this singularity
may be obtained by absorbing
 the $1/M$ correction into a mass renormalization and solving a
self-consistent equation for the mass~\cite{ital}.
 One then obtains $\Pi \sim (k^2 +
m^2)^{-1/2}$, where $m = 2m_0 (1 + O(1/M^{2/3}))$, which is consistent with
 Witten's result for  the mass renormalization.
However, this self-consistent procedure is clearly not exact,
even for large $M$, because near $k^2 =-m^2$, $\delta \sim
M^{-1/3}$, and the
$1/M$ contribution to $\Pi$
has the same order $O(1/M^{-1/3})$ as the leading term. Moreover,
simple estimates
show that the higher-order $1/M$ corrections behave as $\delta^{-1}
(M\delta^3)^{-l}$ and therefore all have the same order near $k^2 = - m^2$.

We will demonstrate in the next section that the self-consistent
procedure~\cite{ital} yielding $\Pi \sim (k^2 + m^2)^{-1/2}$~ is
actually incomplete
because the $1/M$ corrections contribute to both the mass renormalization
and to the renormalization of the spinon wave function. To see this, we will
need to solve the problem exactly by summing up the series of the most
divergent $1/M$ corrections (but we will still assume that $M \gg 1$ and
neglect terms which have relative smallness in $1/M$).

\subsection{Solution of ladder series}

The relevant diagrams  at each order in $1/M$
are presented in Fig~\ref{ladder}. These diagrams form a ladder
series. We will first discuss the solution of the
ladder series and then show that other
diagrams are relatively smaller at large $M$.

It is convenient to formally include both self-energy and vertex
renormalization terms into an effective vertex renormalization, such that
\begin{equation}
\Pi (k)= k_B T~ \int ~\frac{d^2 p}{4\pi^2}
\Gamma({\vec p}, {\vec k})~G(\vec{p} +\vec{k}/2)G(\vec{p} +\vec{k}/2) .
\end{equation}
The integral equation for the vertex can be written down as
\begin{eqnarray}
\Gamma(\vec{p},\vec{k}) = &&1 +  \frac{k_B T}{M} \int
{}~\frac{d^2 l}{4\pi^2} ~
\Gamma(\vec{p} + \vec{l}, \vec{k})~
G(\vec{p} +\vec{k}/2 +\vec{l})G(\vec{p} -\vec{k}/2 +\vec{l})\nonumber\\&&
\times
(2\vec{p} +\vec{k} +\vec{l})_{\mu}~(2\vec{p} -\vec{k} +\vec{l})_{\nu}
{}~D_{\mu\nu}(l) \nonumber\\
&&+  \frac{2 k_B T}{M} \int
{}~\frac{d^2 l}{4\pi^2} ~ \Gamma(\vec{p}, \vec{k})
G(\vec{p} +\vec{k}/2) G(\vec{p} +\vec{k}/2 +\vec{l})\nonumber\\&&\times
(2\vec{p} +\vec{k} +\vec{l})_{\mu}~(2\vec{p} +\vec{k} +\vec{l})_{\nu}
{}~D_{\mu\nu}(l).
\label{B12}
\end{eqnarray}
Again, the angular integration over
both ${\vec p}$ and ${\vec l}$ is confined to a narrow region ($\sim \delta)$
in which both internal
momenta are nearly orthogonal to the external momentum ${\vec k}$. We assume
that $\Gamma ({\vec p}, {\vec k})$ and $\Gamma ({\vec p} + {\vec l}, {\vec k})$
are nonsigular for these values of the angles. We then can set ${\vec p},
{\vec l}$ to be orthogonal to ${\vec k}$ in $\Gamma$ and perform angular
intergation in the Green functions.
 Restricting,
as before, to the integration only near $\pi/2$, and extending the integration
over $p$ and $l$ from $-\infty$ to $+\infty$, we obtain
\begin{equation}
\Pi (k)=\frac{k_B T}{2m_0}~\int_{-\infty}^{\infty}\frac{dp}{2\pi}~
\frac{\Gamma(p,k)}{p^2 +\delta^2} ,
\label{B9}
\end{equation}
where
\begin{equation}
\Gamma(p) = 1 + \frac { m_0 k_B T}{2 \pi M}
{}~\int_{-\infty}^{\infty} dl D(l) \frac{\Gamma(p - l,k)}{(p - l)^2 +\delta^2}
-
\frac { m_0 k_B T}{2 \pi M}~\frac{\Gamma(p,k)}{p^2 +\delta^2}~
\int_{-\infty}^{\infty} dl D(l) .
\label{B99}
\end{equation}
Substituting now  $\Gamma(p,k)=(p^2 +\delta^2)\Psi(p)$, we can rewrite
(\ref{B99}) as
\begin{equation}
(p^2 +\delta^2)\Psi(p)= 1 +  \frac { m_0 k_B T}{2 \pi M}
\int_{-\infty}^{\infty}dl D(l)\left(\Psi(p) - \Psi(p - l)\right).
\end{equation}
Finally, performing a Fourier transformation to real space,
\begin{equation}
\Psi(x)=\int_{-\infty}^{\infty}\frac{dp}{2\pi} e^{ipx} \Psi(p),
\end{equation}
we obtain for $\Psi (x)$ an {\it one-dimensional} Schrodinger equation with a
source
\begin{equation}
\left(-\frac{d^2}{dx^2} + V(x) +\delta^2 \right)\Psi(x)= \delta(x).
\label{B7}
\end{equation}
The potential $V(x)$ is given by
\begin{equation}
V(x)= \frac{m_0 k_B T}{M}\int_{-\infty}^{\infty}\frac{dl}{2\pi}
D(l) \left(1 - e^{-ilx} \right) = \frac{6\pi m_0^3}{M} |x|.
\end{equation}
Notice that the integral over $l$ is free from divergencies.
 This is again the result of including  both
self-energy and vertex corrections in
the ladder series.

Going back to (\ref{B9}),
 we see that $\Pi (k)$ takes a
simple form
\begin{equation}
\Pi(k)=  \frac{k_B T}{2m_0} \Psi(x=0).
\label{B999}
\end{equation}
We emphasize at this point that $\Psi(x)$ is {\it not} a
wave function of the 1D Schrodinger
equation, but rather a Green's function of the inhomogeneous differential
equation (\ref{B7}). This equation was solved
earlier in the content of the weak
localization theory \cite{alt}, and we simply quote the result:
\begin{equation}
\Psi(x=0)= -\frac{1}{2} \left(\frac{6 \pi m_0^3}{M} \right)^{-1/3}
\frac{Ai(s)}{Ai'(s)},
\end{equation}
where $s=\delta^2~(6 \pi m_0^3/M)^{-2/3}$,
 $Ai (s)$ is the Airy function, and
$Ai'$ is its derivative with respect
to the argument. For  positive $s$,
both $Ai$ and $Ai'$ decay exponentially, but for negative $s$, they
have zeros~\cite{abramowitz}. The singular contributions to $\Psi(x=0)$
indeed come from the zeros of $Ai'$. Expanding near each of these zeros,
we obtain
\begin{equation}
Ai'(s)\approx (s - s_n)~Ai''(s_n)= -|s_n| (s - s_n) Ai(s_n)
\end{equation}
where $s_n <0$ is  the n-th zero of
the derivative of the Airy function.
Clearly, the polarization operator now has a set of simple poles at $\delta^2 =
s_n (6\pi m^{3}_0/M)^{2/3}$.

Finally, we remind that we are actually interested in the long-distance
behavior of the spin correlation function.
At  long  distances, only the pole with the smallest mass
is relevant. Near this pole,
 we can approximate
the polarization operator and, hence, the staggered susceptibility as
\begin{equation}
\chi (k,0)=\frac{N^{2}_0}{\varrho_s}~ \frac{k_B T}{2\pi \varrho_s}
{}~\left(\frac{2}{M}\right)^{1/3}~\frac{Z}{k^2 + m^2},
\label{B100}
\end{equation}
where $Z = (3\pi^4)^{1/3}/|s_1|$, $s_1 \simeq - 1.02$ \cite{abramowitz},
and the renormalized
mass $m^2$ is given by
\begin{equation}
m^2 = 4 m_0^2 \left(1 - s_1 \left(\frac{6 \pi}{M} \right)^{2/3} \right) .
\label{B11}
\end{equation}
This last result coincides with the expression for the mass obtained by
Witten~\cite{wit}.

Eq. (\ref{B100}) is the key result of this section. We see that the
series of divergent $1/M$ corrections near the branch cut not only produce
nonanalytical renormalization of the two-spinon mass, but also change the
branch cut behavior of the staggered susceptibility
to a set of poles at discrete values of $k$.
In real space, we then have
$\chi(r) \sim M^{-1/3}~\exp(-m r)/\sqrt{r}$  at very large distances which
agrees with the result for the $n-$field sigma-model.
Notice, however, that the residue
of each pole contains a factor  $M^{-1/3}$ and vanishes in the limit of
$M=\infty$. In this limit, the spacing between neighboring poles vanishes, and
one recovers the mean-field branch-cut solution for $\Pi (k)$, and hence
$\chi(r) \sim \exp(-2m_0 r)/r$. It is not difficult to  show that
at finite but large $M$, the asymptotic behavior associated with the lowest
pole exists only at very large distances, $m r > M^{1/3}$, while at smaller
distances the behavior of spin correlations remains the same as in the
mean-field theory. Indeed, at large $M$, the asymptotic behavior is of no
practical relevance, since at $m r \sim M^{1/3} \gg 1$,
 the spin correlation function
is already negligibly small. However, for the physical case of $M=2$, we expect
the behavior associated with the pole to dominate at all scales larger than the
correlation length.

Notice, however, that the existence of the nonanalytical corrections to the
spin correlation length makes it
very difficult to obtain
the exact expression for $\xi = m^{-1}$
in the $CP^{M-1}$ model. The exact value of $\xi$ in the
 $O(N)$ model is known at arbitrary $N$~\cite{csy,ital2}. The $M=2$
value of the overall factor $Z$ in the staggered susceptibility (\ref{B100})
($Z =6.517$) is also substantially larger
than $Z \sim 2$
expected for the $O(3)$ model from the $1/N$ expansion~\cite{csy}.

 Before concluding this section, we show that the terms not included in the
ladder series do not contribute to the renormalization of the polarization
operator to leading order in $1/M$. The key point is that if we allow the
gauge field propagators to cross each other even once,  the
integral over the internal momentum of the two Green
functions located between the propagators gives zero
because the poles are located in the same
half-plane. To illustrate this, consider the second, "umbrella"-like,
diagram in Fig~\ref{wrong_dia}.
The analytical expression for this diagram is
\begin{eqnarray}
\Pi_{\text{ext}}=&&\frac{16 m^{4}_0 (k_B T)^2}{M^2}\int\int\int
\frac{d^2 p d^2 l d^2 q}{64 \pi^6}
{}~D(l) D(u) ~\frac{\sin^2\psi \sin^2\gamma}{(p^2 +\delta^2)^2 +4m_0^2 p^2
\cos^2\varphi}\nonumber \\
&&\times \frac{1}{((\vec{p} +\vec{q})^2 +\delta^2)^2 +4m_0^2
(p\cos\varphi +q\cos\gamma)}\times
\frac{1}{((\vec{p}+\vec{l})^2+\delta^2) +2 i m_0 (p\cos\varphi + l\cos\psi)}
\nonumber \\
&&\times
\frac{1}{((\vec{p} +\vec{l} +\vec{u})^2 +\delta^2) + 2 i m_0
 (p\cos\varphi + l\cos\psi + u\cos\gamma)} .
\label{B8}
\end{eqnarray}
Here $\varphi, \psi$ and $\gamma$  are the angles between the external momentum
$\vec{k}$ and the internal ones $\vec{p}, \vec{l}$, and $\vec{q}$,
 respectively.
As before,  the relevant
$M^{-2} (k^2 + \delta^2)^{-7/2}$ contribution from this diagram
is confined to the integration over internal momenta which are nearly
orthogonal
to $\vec{k}$. Expanding
the angles around $\pi/2$, rescaling them in the same way as in (\ref{B4}),
 and  shifting the variable
$\psi \rightarrow \psi -\gamma$, we find that the angular integration
in (\ref{B8}) reduces to
\begin{eqnarray}
&& \int_{-\infty}^{\infty}d\gamma
\int_{-\infty}^{\infty}d\varphi \frac{1}{(p^2 +\delta^2)^2 +\varphi^2}
\frac{1}{((p - q)^2 +\delta^2)^2 +(\varphi +\gamma)^2}\nonumber\\
&&\times
\int_{-\infty}^{\infty}d\psi \frac{1}{((p - l)^2 +\delta^2) + i\psi}~
 \frac{1}{((p - l - q)^2 +\delta^2) + i(\varphi +\psi)}.
\end{eqnarray}
The  integration over $\psi$ then
gives zero. The same is true for all other "crossing" diagrams, as well
as for the "rainbow"-like graphs for the self-energy (see
Fig.~\ref{wrong_dia}). Note that this result
does not depend on the dimensionality.

\section{quantum-disordered region}
\label{sec4}

We now perform the same type of analysis for the QD regime.
This is the low-temperature regime at $g >g_c$, when
long-range order in the ground state is destroyed by quantum fluctuations.
The ground state is then a total spin singlet,
and there is a gap, $\Delta$, towards the lowest excited
triplet state with $S=1$.
The temperature  corrections to the staggered susceptibility
in the QD regime are exponentially small,  and we therefore
restrict our analysis
in this Section to $T=0$. At zero temperature frequency becomes a continuous
variable, and perturbative $1/M$ expansion has to be performed in a
three-dimensional spacetime. The
higher dimensionality has already shown up in the computation of the
the polarization
operator at $M=\infty$, which in the QD regime has only a  weak, logarithmic
singularity at $k^2 = - 4m^{2}_0$ ~(see eq. (\ref{B22})).

\subsection{$1/M$ corrections}

The computation of the leading $1/M$ corrections
 proceeds exactly in the same way as in the RC regime.
One can easily verify that the typical
internal momenta in the $3D$ analog of
(\ref{B3}) are of the order of $\delta$ and nearly orthogonal to the external
momentum  $\vec{k}$. This, in turn, implies that all internal momenta
are confined to the plane perpendicular to $\vec{k}$,
which makes the problem effectively $two-dimensional$.

Technically this can be seen as follows. Consider $3D$ vectors $\vec{p}$
and $\vec{l}$. Let them form angles $\varphi$ and $\psi$ with $\vec{k}$,
respectively. Then
$\vec{p} ~\vec{l} =p l(\cos\varphi \cos\psi + \sin\varphi \sin\psi
\cos\theta)$,
where $\theta$ is the azimuthal angle $(0\leq \theta \leq 2 \pi)$.
In the spirit of our approximation, we expand $\varphi$ and $\psi$
around $\pi/2$. Then immediately
$\vec{p} ~\vec{l}= pl(\cos\theta + O(\delta^2))
\approx pl\cos\theta$, which is the scalar product in 2D.

Performing the angular integration in the same way as in the previous section,
we obtain
\begin{equation}
\Pi_{1/M}(k)=-\frac{1}{4\pi^2 M}
 \int \int
\frac{d^2 p d^2 l}{16\pi^4}~\frac{D(l)}{p^2 +\delta^2}
{}~\left(\frac{1}{p^2 +\delta^2} - \frac{1}{(\vec{p} +\vec{l})^2 +\delta^2}
\right) .
\label{C1}
\end{equation}

Evaluating this integral, we find
\begin{equation}
\Pi_{1/M}(k) = - \frac{3 m_0}{4 \pi M \delta^2}~\left(\log{1/\delta} +
C\right) ,
\label{n1}
\end{equation}
where $C$ is a constant. As in the RC regime, the $1/M$ correction to $\Pi$
has a stronger dependence on $\delta$ than is required for a regular
$1/M$ expansion for the two-spinon mass in (\ref{B22}).
 However, contrary to the previous case,
the $(\log \delta)/\delta^2$ corrections come only from the self-energy
term. Vertex corrections  only contribute to a constant $C$ term in (\ref{n1}).
Meanwhile, if we look back  on how the transformation of a  branch cut into
a pole was obtained in the previous section,
we can see that this transformation is primarily due to vertex corrections.
The inclusion of the
self-energy terms only serves to make the Fourier transform of $D(l)$
infrared-finite, and allows the correct mass renormalization.
 By analogy, we can expect that the series of
logarithmic terms will only contribute to the mass renormalization, but the
 mean-field form of the polarization operator will survive.
We performed an explicit computation of
the ladder series of the logarithmic $1/M$ corrections, and
found that with the logarithmic accuracy,
the polarization operator is given by
\begin{eqnarray}
\Pi_{\text{log}}(k)=&&
\frac{1}{8\pi m_0}\int\frac{dp\cdot p}{p^2 +\delta^2}\Big(1 -\frac{12 m_0^2}
{M}\frac{\log m_0/\delta}{p^2 +\delta^2} +
\left(\frac{12 m_0^2}{M}\frac{\log m_0/\delta}{p^2 +\delta^2}\right)^2
+\cdot\cdot\Big) \nonumber\\
&& =\frac{1}{8\pi m_0}\int\frac{dp\cdot p}{p^2 +\delta^2 +
\frac{12m_0^2}{M}\log\frac{m_0}{\delta}} = \frac{1}{16\pi m_0}
\log(\frac{m^{2}_0}{k^2 + {\bar m}^2}),
\label{C2}
\end{eqnarray}
where
\begin{equation}
{\bar m}^2=4m_0^2 \left(1 + \frac{6}{M}\log\frac{m^{2}_0}{m^2 -
m^{2}_0}\right)=
4m_0^2 \left(1 +\frac{6}{M}\log M +O(M)\right) .
\label{C3}
\end{equation}
We see that $\Pi_{\text{log}}(k)$ still has a logarithmic singularity at $k^2
= - {\bar m}^2$.
 However, this result is still incomplete, and, in fact, we have to go beyond
the logarithmic accuracy and include vertex corrections, which are crucial for
the renormalization of the functional dependence of the propagator.
Therefore, we must again consider the full ladder equation for the polarization
operator.

\subsection{The ladder diagrams}

As before, the ladder equation for the polarization operator is
obtained by multiplying the equation  for the vertex function $\Gamma (p)$
(which is a 3D analog of
(\ref{B12})) by $G(\vec{p} +\vec{k}/2)G(\vec{p} -\vec{k}/2)$
and integrating over $3D$ momentum $\vec{p}$. This procedure is completely
equivalent to that in the RC regime, and we present only the result.
Introducing $\Psi (p) = \Gamma(p)/(p^2 +\delta^2)$, we obtain
\begin{equation}
\Psi(p) (p^2 +\delta^2) = 1
 + \frac{m_0}{M}\int
\frac{d^2 \vec{l}}{4\pi^2} D(l) \Psi(\vec{p} +\vec{l})~
 -\frac{m_0}{M}\int \frac{d^2 \vec{l}}{4\pi^2} D(l) \Psi(\vec{p}).
\end{equation}
In real space, this equation is again equivalent to the
 Schrodinger equation with a source
\begin{equation}
\left(- \bigtriangledown^2 + V(r) +\delta^2 \right)\Psi(r)=\delta(r) ,
\end{equation}
where $V(r)$
is the
Fourier transform of the gauge field propagator
\begin{equation}
V(r)= +\frac{m_0}{M}\int\frac{d^2 \vec{l}}{4\pi^2} D(l)\left(
1 - e^{i\vec{l}\vec{r}} \right) = \frac{12 m_0^2}{M} \log|r|.
\end{equation}
The integral over $\vec{l}$ is again  infrared finite.
The dependence on $M$ in this equation can be eliminated by a rescaling
$r\rightarrow\left(\frac{12m_0}{M}\right)^{-1/2}\tilde{r}$, and we obtain
\begin{equation}
\left(-{\tilde{\bigtriangledown}}^2 + \log|\tilde{r}| +\epsilon \right)
\Psi(\tilde{r})=\delta(\tilde{r}),
\label{C4}
\end{equation}
where the derivatives are taken with respect to $\tilde r$, and
we introduced
\begin{equation}
\epsilon=\frac{M}{12m_0^2}\left(\delta^2 - \frac{6m_0^2}{M}\log
\frac{12m_0^2}{M}\right) = \frac{M}{48m_0^2} (k^2 + {\bar m}^2),
\end{equation}
where ${\bar m}$ is given by (\ref{C3}).
Polarization operator is again  related to
$\Psi (r=0)$ via $\Pi(k)= (\pi/2m_0)\Psi(r=0)$.

Surprisingly, to the best of our knowledge,
no exact solution of the 2D Schrodinger equation with
a logarithmic potential (``2D Hydrogen atom")  is known. However, as
$V(r) \sim \log|r|$ is unbounded from above, there exists a
discrete set of
energy levels for the homogeneous Schrodinger equation, and, just as in the RC
regime, $\Psi(0)$ will have an infinite number of simple poles
at some discrete $\epsilon_n \sim O(1)$ :
\begin{equation}
\Psi(0) \propto \sum_n \frac{A_n}{\epsilon -\epsilon_n}=
\frac{48 m_0^2}{M}\sum_n \frac{A_n}{k^2 + {\bar m}^2 -\frac{48 m_0^2}{M} .
\epsilon_n}
\label{CC}
\end{equation}
Here $A_n$ is the (unknown) residue of the n-th pole.
 Using the WKB approximation
it is easy to estimate that at large $n$, $\epsilon_n \sim \log(n)$,
and $A_n \sim n^{-1}$.
Also observe that momentum $k$ in (\ref{CC})
is a vector in $2+1$ space-time dimensions. Hence,
eq.(\ref{CC}) in fact describes the dynamic staggered susceptibility.

Eq. (\ref{CC}) is the key result of this section. We have found that,
just as was the case in the RC region,
the mean-field expression for the staggered susceptibility is also
invalid at finite
$M$ in the QD region, although the corrections to the two-spinon mass and
to the susceptibility have different dependence on the expansion parameter
$1/M$ in the QD case.
 The staggered susceptibility computed to order $1/M$
contains an infinite number of simple poles.
The lowest pole governs
the behavior of the spin correlation function at very large distances.
Performing the Fourier transform of (\ref{CC}), we obtain
for equal-time spin-spin correlator
 $\chi (r) \sim (1/M) \exp (-m r)/r$,
 compared to the mean-field expression $\chi (r) \sim \exp (- 2 m_0 r)/r^2$.
  As in the RC regime,
there is a crossover between the two regimes, which occurs at $m r \sim M$.

A confinement of spinons
due to the logarithmic potential was qualitatively discussed by Wen~\cite{wen}.
A complimentary scenario of spinon
confinement in the QD phase was considered by Polyakov~\cite{pol2}, and
more recently by Read and Sachdev \cite{rs} and Murthy and Sachdev~\cite{ms}.
They argued that the instanton tunneling events
(which we do not
consider) lead to a {\it linear} confining potential between spinons
at distances $\xi_C \sim {\xi}^{M\cdot \rho_1}$, $\rho_1 \approx
0.06$~\cite{ms}. At large $M$, this scale
is much larger than
the typical confinement  scale, $\sim M \xi$,
which appears in our consideration. The two scales, however, may become
comparable at small $M$.

\section{quantum-critical regime}
\label{sec5}

Finally, we consider the QC regime, which is the intermediate
regime between the RC and the QD regimes.
At $T=0$, the $CP^{M-1}$ model action
describes D=3 critical theory which possess no confinement~\cite{no_conf}.
The $M=\infty$ susceptibility
behaves as $1/k$, and the $1/M$ corrections from
both constraint and gauge fluctuations are of the form
$\log(1/k)/k$. Evaluating the $1/M$ terms and exponentiating the result
we obtain $\chi(k) \sim k^{-(2 - \eta)}$, where
$\eta=1 - \frac{32}{\pi^2 M}$. Notice that the correction to the mean-field
result for $\eta$ is large. On the contrary, for $O(N)$ sigma-model at the
critical point we have $\eta =0$ which is much closer to the Monte-Carlo result
for the $O(3)$ model - $\eta = 0.03$~\cite{holm}.

At finite temperature, the confinement does exist, and
 we proceed in exactly the same way as in the two previous
sections. In the QC regime, the typical frequencies are of the order of $k_B
T$,
so, in principle, one has to perform full frequency summation in the $1/M$
formulas for the polarization operator. However, the situation is
 simpler, as our earlier results show that the confining potential is much
stronger in the RC regime than in the QD regime. As a result, the most
divergent contributions in the $1/M$ series always come from the
$\omega =0$ terms in the
frequency summations. This observation
 makes the analysis of the confinement in the QC regime
very similar to that in the RC regime; the only difference is in the form of
the gauge field propagator. Using (\ref{gauge}) and the results of
Sec.~\ref{sec3}, we obtain after simple manipulations that the
branch-cut behavior of the spin susceptibility
is indeed replaced by a set of poles. Near the lowest pole,
the polarization operator behaves as
\begin{equation}
\Pi_L(k)=\frac{1}{\Theta |s_1|} \left(\frac{6\pi \Upsilon}{M}\right)^{1/3}
\frac{k_B T}{k^2 + {\tilde m}^2},
\end{equation}
where $\Upsilon=(2\Theta^2)/\sqrt 5=0.828471$, and the bound state mass
 is given by
\begin{equation}
\frac{{\tilde m}^2}{(k_B T)^2}=4 \Theta^2 \left(1 - s_1 \left(\frac{6\pi
 \Upsilon}{M}\right)^
{2/3}\right).
\end{equation}

\section{conclusions}
\label{sec6}
We summarize the key results of the paper.
We considered here the form of the static staggered susceptibility in the
$CP^{M-1}$ model of an $M-$ component $complex$ unit vector in two spatial
dimensions. The $M=2$ limit of this model describes
 two-dimensional Heisenberg
antiferromagnet, and the $1/M$ expansion we discuss in this paper is a
systematic way to calculate various observables.
The elementary excitations in the $CP^{M-1}$ model are
$S=1/2$ bosonic spinons. The spin variables are bilinear in spinon fields,
and all physical excitations are collective modes of two spinons.
 In the mean-field approximations, the spinons behave as
noninteracting particles, and the static staggered spin susceptibility has a
branch cut along the imaginary $k$ axis, terminated at $k^2 = - 4m^2_0$,
 where $m_0$ is the mass of a spinon.

We have shown that in all three scaling regimes at finite $T$,
the fluctuations of the gauge field give rise
to divergent $1/M$ corrections near the branch cut. We selected the most
divergent corrections and have shown explicitly that they form ladder series.
We then found that the ladder problem is equivalent to
a Schrodinger equation with the $\delta-$functional source in
 dimension $D-1$,
 where $D$ is the spacetime dimension for the original problem.
Schrodinger equation has a discrete set of solutions, and expanding near each
of the solutions, we obtained, instead of a branch cut,  a sequence of simple
poles. This in turn quantitatively
changes the behavior of the spin correlation function at
large distances compared to the mean-field predictions.

The result that the staggered susceptibility has a pole, and not
a branch cut,  is consistent
with the results of the alternative perturbative technique
for 2D antiferromagnets, which is the $1/N$ expansion for the $O(N)$
 sigma-model for an $N-$component $real$ unit field.
The latter model describes Heisenberg antiferromagnet at $N=3$.
Elementary excitations in the $O(N)$ sigma-model
carry $S=1$, and, at any $N$, the static staggered
spin susceptibility
has a well defined single pole at imaginary momentum $k
= \pm i \xi^{-1}$, where $\xi$ is the correlation length in the system.

Comparing now for the $CP^{M-1}$ and the $O(N)$ sigma-models,
  we conclude that although the differences
between them found at the mean-field level are now gone, in practical
calculations it is more convenient to use the $O(N)$ model.
The reason is that the perturbative series in $1/N$ is regular for this
model, and converges much better than the $1/M$ series for the $CP^{M-1}$
model. The latter can even be nonanalytical in some cases, e.g. when
calculating the spin correlation length.

Our results still leave a  minor
discrepancy between the two approaches: in the $O(N)$
model, the static mean-field susceptibility has a $single$ pole,
 and $1/N$ corrections do
not lead to the appearance of new singularities.
 On the contrary,  in the
$CP^{M-1}$ model, we found, to
leading order in $1/M$, an infinite set of poles.
Indeed, $O(N)$ and $CP^{M-1}$ models are equivalent $only$ for $N=3$ and
$M=2$, when they both describe Heisenberg antiferromagnet.
However, Bethe-ansatz
solution for the $O(3)$ model also indicates that there is
a single mass in the problem because the
free energy as a function of the uniform magnetic field, $h$,
 has a single threshold at $h_c =m$~\cite{pasha,hasen}.
We therefore expect that all the poles, that we have found
in the $CP^{M-1}$ model, except for the lowest one,
should disappear for $M=2$. At present, we can only speculate how
this may occur - the most likely possibility, in our opinion, is that
the solutions of the Schrodinger equation with $n>0$ simply
acquire a finite lifetime
due to higher-order corrections in $1/M$. If this is true, then
the susceptibility in the $CP^{M-1}$ model has a single stable pole
at $any$ $M$, while all other poles that we have found to first order
in $1/M$, form an incoherent continuum
 where both real and imaginary part of susceptibility
are present. This is corroborated by an observation that the
 $n$-field propagator of the $O(3)$ model, evaluated at zero frequency and
along imaginary $k$ axis, also has an
imaginary part for $|\tilde k| > 3 m$ ($\tilde k = ik$) due to
decay of a $S=1$ quanta into three others (the simplest way to see this is
to substitute a fixed-length constraint on $\vec n$ field
by the $u ({\vec n}^2)^2$ interaction term
 and do a perturbative expansion in $u$~\cite{subir}).
We, however,  emphasize that this issue is irrelevant for
 the long-distance behavior of the spin
correlation function as this behavior is  associated
with the lowest-energy ($n=0$) pole, which is always stable.

A final remark. In this paper, we considered a field-theoretic description
of the $commensurate$ antiferromagnet. One can also study the
anisotropic version of the $CP^{M-1}$ model,
with the extra factor $(1 -\gamma)$ in front of $|\bar{z} \partial_{\mu}z|^2$
in (\ref{zzz}) ($0 \leq \gamma \leq 1$).
 For $\gamma \neq 0$ the gauge-field propagator acquires a mass,
which prevents the effective potential $V(r)$ in the Schrodinger equation
from becoming arbitrary large. As a result, even without damping,
the number of poles in the staggered susceptibility decreases and
finally, when  $\gamma$ exceeds some critical value,
the disordered phase possesses deconfined $S=1/2$ bosonic spinons~\cite{ital2}.
The disordered phase with deconfined spinons
was also obtained in ~\cite{css} by a direct $1/M$ expansion
at small $1 -\gamma$. This limit is of  particular interest as the
$z-$ field model with $\gamma \approx 1$ describes
 incommensurate quantum antiferromagnet near the critical point.

\section{acknowledgements}

It is our pleasure to thank D. Frenkel, G. Reiter,
 S. Sachdev, C.S.Ting and P. Wiegmann for useful
discussions.  The work was supported by the University of Wisconsin-Madison
Graduate School (A.C.) and by a grant from the State of Texas (O.S.).

\begin{figure} \caption {The first-order $1/M$ corrections
to the polarization operator of spinons.}
\label{f_o_dia}
\end{figure}

\begin{figure} \caption {The diagrammatic representation of the spinon and
the gauge field propagators and of the spinon-gauge field interaction vertex.}
\label{vertices}
\end{figure}

\begin{figure} \caption {The full polarization operator and the
equation for the full vertex function.
The most divergent corrections to the mean-field value of $\Pi (k)$
near $k = \pm 2 i m_0$ comes form ladder series.
 For convenience, both vertex and self-energy terms are
absorbed into the vertex renormalization.}
\label{ladder}
\end{figure}

\begin{figure} \caption {Examples of the
second-order diagrams which do not contribute
to the ladder series. All these diagrams are less divergent near the branch cut
singularity than those included in Fig.~\protect\ref{ladder}}
\label{wrong_dia}
\end{figure}


\begin{references}

\bibitem[*]{oas} On leave from the Institute for High Pressure
 Physics, 142092, Troitsk, Moscow Region, Russia

\bibitem{chakr} S.Chakravarty, B.I.Halperin, and D.R.Nelson, \prb {\bf 39},
 2344 (1989)

\bibitem{csy} A.V.Chubukov, S.Sachdev, and J.Ye, \prb {\bf 49}, 11919 (1994)

\bibitem{css} A.V.Chubukov, S.Sachdev, and T.Senthil, Nucl.Phys.B {\bf 426},
 601 (1994)

\bibitem{pasha} P.B. Wiegmann, Phys. Lett. B {\bf 152}, 209 (1985);
see also A. Polyakov and P.B. Wiegmann, Phys. Lett. B {\bf 131}, 121 (1983)

\bibitem{hasen} P.Hasenfratz and F.Niedermayer, Phys.Lett.B {\bf 268}, 231
 (1991);\\
 P.Hasenfratz, M.Maggiore, and F.Niedermayer, Phys.Lett.B {\bf 245},
 522 (1990)

\bibitem{pol} A.M.Polyakov, $Gauge$ $Fields$ $and$ $Strings$ (Harwood,
 New York, 1987)

\bibitem{Brezin} E. Brezin and J. Zinn-Justin, \prb {\bf 14}, 3110 (1976);
V.L. Pokrovskii, Adv. Phys., {\bf 28}, 595 (1979).

\bibitem{aa} D.P.Arovas and A.Auerbach, \prb {\bf 38}, 316 (1988)

\bibitem{chakrev} S. Chakravarty in {\it{High-Temperature Superconductivity}},
edited by K. Bedell, D. Coffey, D.E. Meltzer, D. Pines and J.R. Schrieffer
(Addison-Wesley, Reading, MA) p.136 (1990).

\bibitem{ital} M.Campostrini and P.Rossi, \prd {\bf 45}, 5621 (1992)

\bibitem{me2} O.A.Starykh, \prb {\bf 50}, 16428 (1994)

\bibitem{wit} E.Witten, Nucl.Phys.B {\bf 149}, 285 (1979)

\bibitem{ital2} M.Campostrini and P.Rossi, Rivista Del Nuovo Cimento
 {\bf 16}, 1 (1993)

\bibitem{adda} A.D.Adda, P.Di Vecchia, and M.Luscher, Nucl.Phys.B {\bf 146},
 63 (1978)

\bibitem{rs} N.Read and S.Sachdev, \prb {\bf 42}, 4568 (1990)

\bibitem{comm} The contribution to
$\Pi_{1/M}$ of the same order of magnitude as in (\protect\ref{B5})
($\Pi_{1/M} \sim \delta^{-4}$)
could, in principle, also come from the
region in the momentum space where $p,l =O(\delta^2)$,
while the values of the angles are arbitrary.
 However, we have verified by direct calculations
that the net contribution
to $\Pi_{1/M}$ from this region
is zero to order  $\delta^{-4}$.

\bibitem{alt} B.L.Altshuler, A.G.Aronov, and D.E.Khmelnitsky, J.Phys.C:Solid
 State Phys., {\bf 15}, 7367 (1982)

\bibitem{abramowitz} M.Abramowitz and I.Stegun, $Handbook$ $of$ $Mathematical$
 $Functions$ (Dover, New York, 1970)

\bibitem{wen} X.G.Wen, \prb {\bf 39}, 7223 (1989)

\bibitem{pol2} A.M. Polyakov, Nucl. Phys., {\bf B120}, 429 (1977).

\bibitem{ms} G.Murthy and S.Sachdev, Nucl.Phys.B {\bf 344}, 557 (1990)

\bibitem{no_conf} T.Appelquist and R.Pisarki, \prd {\bf 23}, 2305 (1981)

\bibitem{holm} P.Peczak, A.M. Ferrenberg and D.P. Landau, \prb {\bf 43}, 6087
(1991)

\bibitem{subir} We thank S. Sachdev for bringing this point to our attention.

\end{references}
\end{document}